\documentclass[]{aa}

\usepackage{graphicx}
\usepackage{txfonts}
\begin{document}

\title{Comparison of the sidereal angular velocity of subphotospheric layers and
       small bright coronal structures during the declining phase of solar cycle 23}
\author{A.~Zaatri \inst{1,}\inst{2}
\and  H.~W\"ohl \inst{1}
\and M. Roth\inst{1}
\and T.~Corbard\inst{2}
\and R.~Braj\v sa\inst{3}}
\institute{Kiepenheuer-Institut f\"ur Sonnenphysik, Sch\"oneckstr. 6, 79104 Freiburg, Germany \\
email: amel@kis.uni-freiburg.de, hw@kis.uni-freiburg.de, mroth@kis.uni-freiburg.de
\and Universit\'e de Nice Sophia-Antipolis, CNRS, Observatoire de la C\^ote d'Azur, BP 4229, 06304 Nice CEDEX 4, France \\
email: corbard@oca.eu
\and Hvar Observatory, Faculty of Geodesy, University of Zagreb, Ka\v ci\'ceva 26, 10000 Zagreb, Croatia \\
email: romanb@geof.hr}
\titlerunning{Solar differential rotation comparisons}
\authorrunning{Zaatri et al.}
\date{Received date / Accepted date}

\abstract
{We compare solar differential rotation of subphotospheric layers derived from local helioseismology analysis
of GONG++ dopplergrams and the one derived from tracing small bright coronal structures (SBCS)
using EIT/SOHO images for the period August 2001 -- December 2006, which correspond to the declining phase
of solar cycle 23.}
{The study aims to find a relationship between the rotation of the SBCS and
the subphotospheric angular velocity. The north-south asymmetries of both rotation velocity measurements are also investigated.}
{Subphotospheric differential rotation was derived using ring-diagram analysis of GONG++ full-disk dopplergrams of
1~min cadence. The coronal rotation was derived by using an automatic method to identify and track the small bright 
coronal structures in EIT full-disk images of 6 hours cadence.}
{We find that the SBCS rotate faster than the considered upper subphotospheric layer (3Mm) by about 0.5~deg/day at the 
equator. This result joins the results of several other magnetic features (sunspots, plages, faculae, etc.) with a higher rotation than the solar plasma. The rotation rate latitudinal gradients of the SBCS and the subphotospheric layers are very similar.
The SBCS motion shows an acceleration of about $0.005^{\circ}$day$~^{-1}$/month during the 
declining phase of solar cycle 23, whereas the angular velocity of subsurface layers does not display any evident 
variation with time, except for the well known torsional oscillation pattern. Finally, both subphotospheric and coronal rotations of the southern hemisphere are predominantly larger than those of the northern hemisphere. At latitudes where the north-south asymmetry of the angular velocity increases (decreases) with activity for the SBCS, it decreases (increases) for subphotospheric layers.}
{}

\maketitle

\begin{table*}
\caption{Unsigned differential rotation parameter {\it B} derived from the linear fit of the rotation averaged over 3 
time frames [Aug 2001 -- Dec 2006; Aug 2001 -- Mar 2004; Apr 2004 -- Dec 2006] and sin$^2(\theta)$ for all latitudes [$-45^{\circ}$,$45^{\circ}$], southern hemisphere [$-45^{\circ}$,$0^{\circ}$] and northern hemisphere [$0^{\circ}$,$45^{\circ}$].}
\label{table_B}
\centering
  \begin{tabular}{ c c c c c c c c c c}
    \hline\hline
    \multicolumn{10}{}{~~} \\
    \multicolumn{4}{c}{August 2001-December 2006} & \multicolumn{3}{c}{August 2001-March 2004} & \multicolumn{3}{c}{ April 2004-December 2006}\\
    \multicolumn{10}{}{~~} \\
    \hline
    \multicolumn{10}{}{~~} \\
                   & All latitudes  & North & South & All latitudes & North & South& All latitudes  & North & South\\
    \multicolumn{10}{}{~~} \\
    SBCS    & $2.72\pm0.04$ & $2.77\pm0.04$ & $2.64\pm0.04$  & $2.70\pm0.03$& $2.70\pm0.04$ & $2.69\pm0.04$& $2.73\pm0.07$& $2.82\pm0.07$ &$2.59\pm0.08$ \\
    $3$~Mm  & $2.77\pm0.01$ & $2.74\pm0.01$ &  $2.73\pm0.01$ & $2.77\pm0.01$& $2.77\pm0.02$ &$2.70\pm0.02$ & $2.77\pm0.01$& $2.72\pm0.02$ &$2.76\pm0.02$ \\
    $6$~Mm  & $2.78\pm0.01$ & $2.77\pm0.01$ &  $2.73\pm0.01$ & $2.78\pm0.01$& $2.80\pm0.02$ & $2.70\pm0.02$& $2.78\pm0.01$& $2.74\pm0.02$ &$2.76\pm0.02$ \\
    $15$~Mm & $2.88\pm0.01$ & $2.90\pm0.01$ &  $2.80\pm0.01$ & $2.88\pm0.01$& $2.93\pm0.02$ & $2.77\pm0.02$& $2.88\pm0.01$& $2.87\pm0.02$ &$2.82\pm0.02$ \\
    \multicolumn{10}{}{~~} \\
    \hline
  \end{tabular}
\end{table*}

\section{Introduction}
The solar angular rotation velocity is a function of latitude, time, and height above or depth below the solar photosphere.
This phenomenon is known as the solar differential rotation. The most commonly established latitudinal dependence of the angular velocity
is given by the empirical relation $\Omega(\theta)=A+B$sin$^2\theta+C$sin$^4\theta$,
where $\theta$ is the latitude, $A$ is the equatorial rotation, $B$ and $C$ are the differential rotation
coefficients, and $C$ is usually neglected for low and mid-latitude measurements. In terms of amplitude,
the rotation of the quiet Sun regions, given for instance by Doppler-shift measurements
of Fraunhofer lines, is found to be slower than the rotation of the photospheric magnetic tracers 
(individual sunspots, sunspot groups, faculae, supergranules, etc.). Moreover, magnetic features exhibit different 
angular velocities depending on their evolution, age, and size (see a review by Beck, 2000). So far, this difference is explained 
by magnetic features being rooted at different depths (Ru\v zdjak et al., 2004). Several local and global helioseismology analyses of 
solar acoustic modes, such as time-distance helioseismology analysis (e.g., D'Silva, 1996) and f-mode analysis 
(e.g., Corbard \& Thompson, 2002), have confirmed that the angular velocity increases with depth in the upper layers 
of the convection zone, at intermediate latitudes.

The coronal angular velocity has been measured using several coronal structures such as coronal green lines 
(e.g, Altrock, 2003; Badalyan et al. 2006), radio emission flux (e.g, Mouradian  et al., 2002), and coronal holes 
(e.g, Insley et al., 1995). Many of these features show two rotational modes where high-latitude regions rotate 
more rigidly than low-latitude regions. However, part of this behaviour has been related to the lifetime of the structure
and not specifically to that of the corona (see a review by Schr\"oter, 1985). Moreover, small bright coronal structures 
(SBCS, hereafter) have been used to estimate the coronal rotation from data recorded by the EUV imaging telescope (EIT) 
on board the solar and heliospheric observatory (SOHO). These features are mostly short-lived structures with a lifetime up to 54 hours (Braj\v sa et al., 2008) and are formed at a height of about 8000-12000 km above the photosphere (Braj\v sa et al., 2004). 
Braj\v sa et al. (2001,2002) explored these structures using 284 \AA~EIT filtergrams with a 6 hr cadence, whereas Karachik et al. (2006) 
used 194 \AA~EIT filtergrams with 12~min cadence. Both authors confirm the differential rotation
of the corona through the SBCS with a close relation to the differential rotation of atmospheric magnetic features. 
Moreover, the SBCS are interesting features to be used for the rotation estimation 
along the cycle since they appear at both minimum and maximum cycle phases, in contrast to other activity features that are absent at minimum activity or, for instance, coronal holes with their polar concentration at that period. Their relation to the photospheric magnetic field has been studied by
Pres and Phillips (1999), who find that the time evolution of bright coronal points observed with EIT/SOHO is 
correlated with the rise and fall of the magnetic field given by the MDI/SOHO magnetograms.

In this paper, we compare the rotation velocity of subphotospheric layers (from 3Mm to 15Mm) obtained from ring-diagram analysis of GONG++ data using the GONG ring-diagram analysis pipeline\footnote{http://gong2.nso.edu/archive/patch.pl?menutype=h} (Corbard et al.,  2003) with the one measured by tracing the SBCS observed in EIT/SOHO images (at 284 \AA) using the automatic method described in Braj\v sa et al. (2001). The purpose of this comparison is to get a complementary view of the relation between coronal features and their root layers below the photosphere. Moreover, we investigate the latitudinal gradient of the angular velocity to check the assumption of the two rotational modes and the 
variation of the latitudinal profile of the coronal rotation during the declining phase of the solar cycle 23 as reported 
by Altrock (2003). Finally, a comparison between the rotations of the northern and the southern hemispheres is 
shown for both SBCS and subphotospheric layers.

\section{Data reduction}

\subsection{Rotation of subphotospheric layers}
Subphotospheric angular velocity is measured using ring-diagram analysis which is a local helioseismology technique based
on frequency-shift measurements of high-degree acoustic-modes to infer horizontal velocity flows at different subphotospheric
depths (Hill, 1988). By dividing the full-disk image into 189 overlapping patches with centres separated by $7.5^{\circ}$ 
in latitude and longitude with a latitude range of [-52.5,52.5] and a central meridian distance range of the same extent, 
each region is remapped and tracked at the surface rotation rate of its central latitude in consecutive images with a 1~min
cadence. The resulting data cubes are Fourier-transformed, and the resulting power spectra are fitted with a Lorentzian
profile to derive frequency shifts that are directly related to horizontal velocity components by assuming a plane wave
approximation. The horizontal velocity as function of depth is then deduced using inversion methods. 

We considered the depth range [3Mm,15Mm]. We avoided extending the radial depth to upper layers since surface measurements of the flow are affected
by poor estimation of the inversion kernels because of the complexity of the photospheric layers. Deeper
measurements were not possible because the small size of the patches only gives access to high spatial frequencies (high spherical harmonic degree). The sidereal angular velocity for each local area (latitude, Carrington longitude) was recovered by adding the measured east-west flow (zonal 
flow) to the tracking rate. Finally, the rotation rate at each latitude is given by averaging over all the available 
Carrington longitudes for each month. The synodic-sidereal correction was taken into account for each day.

\subsection{Rotation of small bright coronal structures}
The data reduction to determine the motions of
bright coronal structures was described in detail by Braj\v sa et al. (2001).
Although not just coronal bright points (CBP) were selected in the
cited paper, this abbreviation was used. In fact, besides points also small loops and small active regions were
selected. We prefer to call the used tracers small bright coronal
structures (SBCS) now.
The interactive method has up to now only been applied to images obtained in 1998
and 1999, because it is very time consuming. An automatic method was developed later to cover more images from
most of the years of an activity cycle. The main idea was to automatically select small structures in three consecutive
images and find whether they belong to a stable structure within
12 hours so they only show small changes in latitude and
in longitude. Slight differences were applied to the parameters of the
automatic reduction program (Braj\v sa et al. 2001).
First, the circumferences of the allowed structures was changed to the range
of 30 to 80 pixels, which equals 80 to 210 arcseconds. The range
of relative intensities was 100 to 600 units.
Most important was a change in the allowed difference in latitude
from one image to the next from 1 degree to 0.8 degree and the application
of a differential rotation for the allowed motions in longitude:
The limits were 9.5 to 16.5 degrees per day of the  synodic rotation,
but these limits were reduced by $- 3.0 * $sin$^2 (\theta)$, where $\theta$
is the latitude. 

The conversion of synodic rotation rates to sidereal was performed for each observing
time separately. Filtering after a first fit to the differential rotation law with $C=0$ was applied
in the same way as described in Section 4.2 of Braj\v sa et al. (2001).
This filtering was applied only when data from certain periods (e.g.
months, years or longer periods) are combined. Thus the sum of
all structures used is slightly different when adding monthly numbers
and comparing with annual sums.
At the beginning the data were selected for all latitudes in each
month and only differential rotation parameters fitted using $C=0$.
To adapt the reduction to the GONG++ data, a new program was
added, where the results were sampled for overlapping bins of
15 degrees width in latitude and a stepping of 7.5 degrees.
Thus a data cube was derived with rotation rate results and
their errors for 15 latitude bins and 65 months
from August 2001 until December 2006.

\begin{figure}
\centering
\small
\includegraphics[width=9cm]{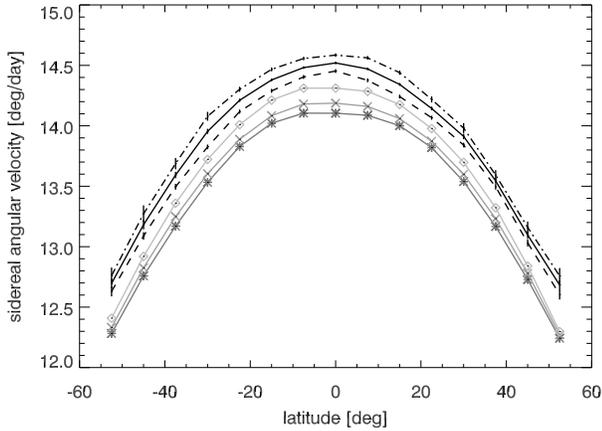}
\caption{Rotation velocity derived from ring-diagram analysis at 3Mm (asterisk), 7Mm (cross), 15Mm (diamond) (lines from dark to light grey) and SBCS 
tracing (black) averaged over the periods: August 2001 -- March 2004 (dashed), April 2004 -- December 2006 (dashed dotted).}
\label{velocity}
\end{figure}

\begin{figure}
\small
\centering
\vbox{
\includegraphics[width=9cm]{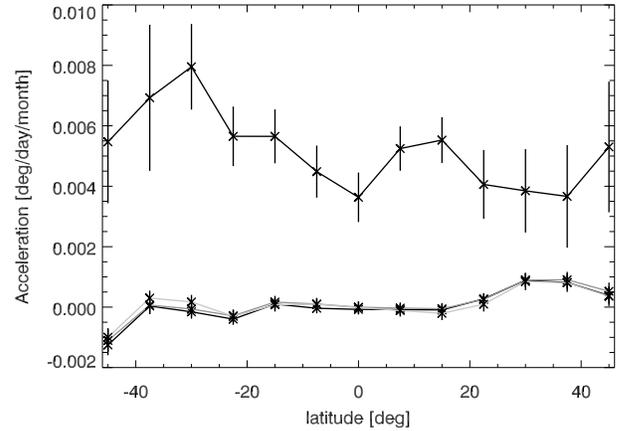}
\includegraphics[width=9cm]{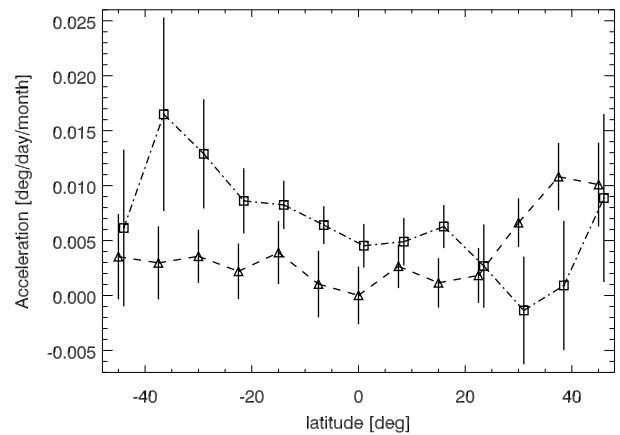}
}
\caption{Upper panel: Slope of the linear fit of the rotation with time derived from ring analysis at 3Mm, 6Mm and 15Mm 
(from dark to light grey) and SBCS tracing (black) for the period Aug 2001 -- Dec 2006. 
Lower panel: Slope of the linear fit of the SBCS angular velocity with time for the periods: Aug 2001 -- Mar 2004 (dashed line, triangles); Apr 2004 -- Dec 2006 (dashed-dotted line, squares). Squares are shifted by $1^{\circ}$ to avoid overlapping of error bars.}
\label{slopes}
\end{figure}

\section{Solar differential rotation}
We measured the subphotospheric and coronal rotation velocities for the period August 2001 -- December 2006, which corresponds to the declining phase of solar cycle 23. GONG++ data covers all the months with a good duty cycle. 
The variation in the zonal flow (i.e rotation residual) as derived from ring-diagram analysis of GONG++ data during this period has been studied in detail by Komm et al. (2009) with 3 extra months in 2007. They find a positive correlation between the unsigned magnetic flux and the zonal flow and report a higher zonal flow in patches with activity than for the quiet Sun patches. 
They do not find any particular long-term pattern of the zonal flow except the well-known pattern of torsional oscillations
(Howard and LaBonte, 1980). However, a slight increase in the zonal flow in patches with activity was observed at
minimum activity. 
Figure \ref{velocity} shows the sidereal angular velocity given by the coronal and subphtospheric measurements for the latitudinal range 
[$-52.5^{\circ}$,$52.5^{\circ}$]. From ring-diagram analysis measurements, we see that deep subsurface layers rotate faster than those closer to the surface with a 
decreasing radial gradient as the latitude increases. The radial gradient variation in the rotation velocity estimated 
from ring-diagram analysis of the same data was studied in detail by Zaatri and Corbard (2009), who report that the 
use of higher latitudes can lead to a reversal of the radial gradient from negative to positive. Moreover, 
Fig. \ref{velocity} shows a smaller northern radial gradient than the southern gradient at high latitudes.
From SBCS tracing, the rotation rate is found to be higher than for all the considered subsurface layers by about 
$0.5^{\circ}$day$^{-1}$ at the equator (this corresponds to a velocity of about 70~m/s, and a frequency of about 16~nHz) 
compared to that of the subsurface layer at 3Mm depth. This difference is in good agreement with the angular velocity 
difference between the photospheric active regions and their surrounding quiet solar plasma (e.g, Koch, 1984). 

The upper panel of Fig. \ref{slopes} shows the angular acceleration of the SBCS and of the subsurface layers. 
The subphotospheric layers have no significant acceleration compared to that of the SBCS. However, it is worth mentioning that the absence of acceleration at low latitudes and the week acceleration at latitudes higher than $20^{\circ}$ (up to $10^{-3}$ deg/day/month) is consistent with the torsional oscillation pattern for the years 2001 -- 2006 (see Fig. 30 of Howe 2009).
 The SBCS rotation shows a positive increase with an acceleration of about $0.005^{\circ}$day$^{-1}$/month. This acceleration also depends on activity as shown by the lower panel of Fig. \ref{slopes}. 
At higher activity, the acceleration is clearly seen to be lower than the lower activity for the southern hemisphere; 
however, it shows a more complicated latitudinal and activity variation in the northern hemisphere. It is worth mentioning 
that the photospheric magnetic features are also rotating faster at low activity, as seen from several observations 
(e.g, Braj\v sa et al., 2006). Lastly, we investigated the latitudinal gradient of the rotation by measuring the 
differential rotation coefficient $B$. We avoided taking the highest latitude $\pm52.5^{\circ}$ where the SBCS are almost absent 
at low activity, leading to angular velocities which are mostly interpolated values. 
Table 1 gives the $B$ values for both SBCS and three subphotospheric layers, and $B$ is seen to increase with depth at 
the outer part of the convection zone. Moreover, neither the small SBCS nor the subphotospheric 
layers show any significant variation with the activity of the latitudinal gradient of their angular velocity.

\section{North-south asymmetry} 
\begin{figure}
\centerline{
\small
\vbox{
\includegraphics[width=9cm]{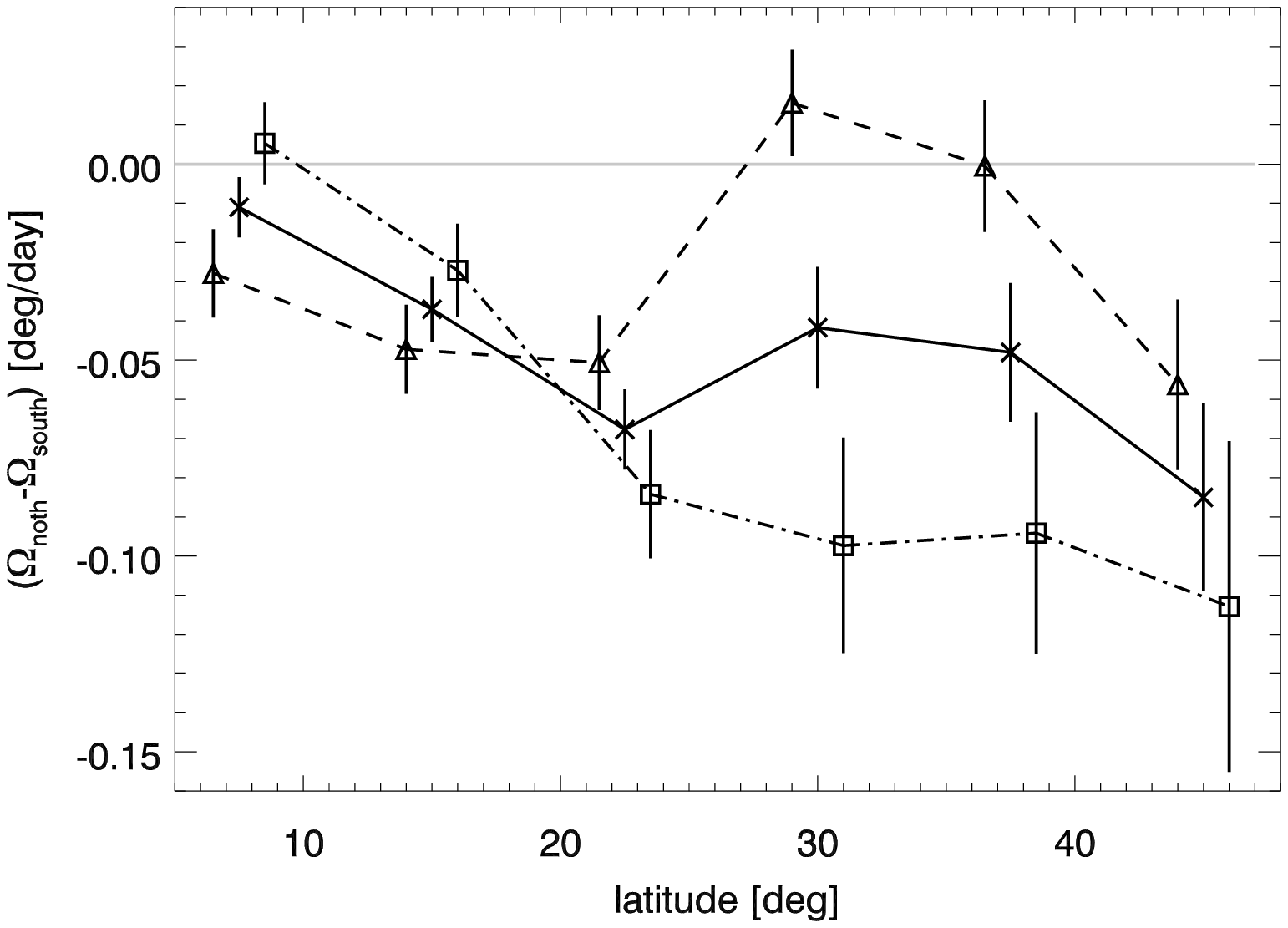}
\includegraphics[width=9cm]{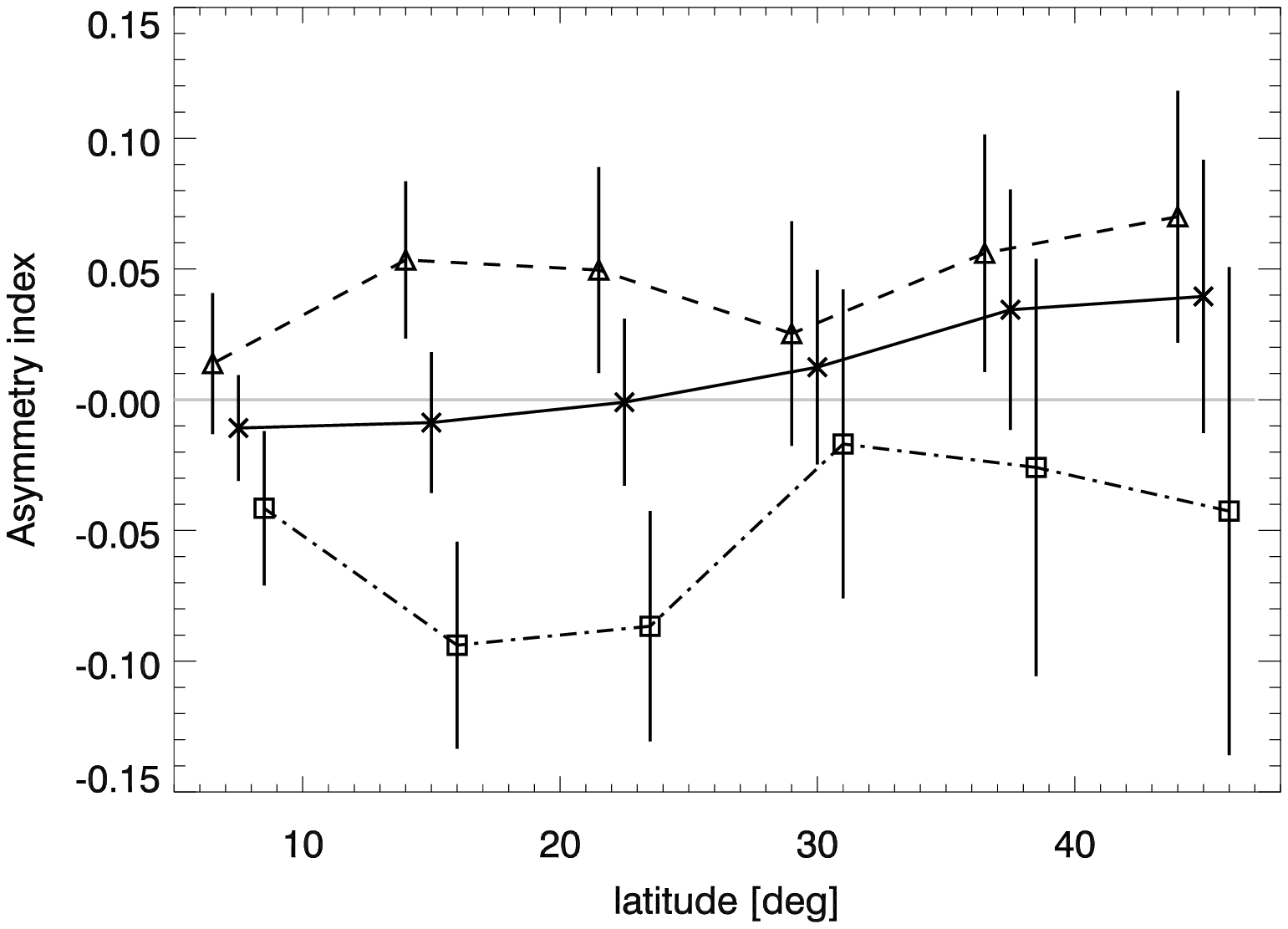}
}
}
\caption{Upper panel: Difference between the northern and southern angular velocity of SBCS averaged over the periods August 2001 -- December 2006 (full lines), August 2001 -- March 2004 (dashed,triangles), April 2004 -- December 2006 (dashed-dotted,squares). Lower panel: Asymmetry index of the coronal activity derived from the number of observed SBCS in the north (N) and in the south (S) given by the formula (N-S)/(N+S) by summing the structures over the periods given in the upper panel. The thin grey line shows the zero level. Triangles and squares are shifted by $1^{\circ}$ to the left and right to avoid overlapping the error bars.}
\label{n-s_eit}
\end{figure}

We investigated the north-south asymmetry of the coronal rotation velocity through the SBCS angular velocity change between the northern and southern hemispheres by evaluating the difference between two symmetric bins. For instance, a rotation velocity averaged over a bin centred at $15^{\circ}$ is subtracted from that derived from a bin centred at $-15^{\circ}$, the difference value is shown at latitude $15^{\circ}$. This quantity is evaluated for the unsigned latitude range [$7.5^{\circ}$,$45^{\circ}$] in the upper panel of Fig. \ref{n-s_eit}. The north-south asymmetry of the coronal rotation reveals a faster southern hemisphere during the period under analysis. Also shown is a latitudinal dependence of this property with an increasing north-south difference in rotation from the equator to about $20^{\circ}$, where the amplitude is more pronounced with strong activity. At latitudes higher than $20^{\circ}$, this asymmetry continues to increase at the low activity level (end of cycle 23), whereas it is more irregular at the beginning of the declining phase and becomes even positive around $30^{\circ}$. To see whether the asymmetry in the coronal rotation velocity is related to the coronal activity, we investigated the activity asymmetry via the asymmetry index given by $(N-S)/(N+S)$ for each pair of symmetric bins where $N$ is the number of structures in the northern bin and $S$ is the number of structures in the southern bin. The asymmetry index and its statistical mean error are evaluated on a monthly basis and shown in Fig. \ref{n-s_eit}. The figure shows that the northern coronal activity prevails during the first half of the declining phase of solar cycle 23 and the southern activity is more dominant during the second half of this period. This agrees with the flare activity during this cycle, which favoured the north at maximum activity and the south during the declining phase (Joshi et al. 2007). Figure \ref{n-s_eit} clearly shows that, in the first half of the declining phase of solar cycle 23, the northern hemisphere, which is more active, rotates more slowly than the southern hemisphere, and it remains slower at the end of the cycle even though it becomes less active than the southern hemisphere. This disagrees with several studies reporting that the active hemisphere rotates more slowly than the less active one (Obridko et al. 2001). 

\begin{figure}
\centerline{
\small
\vbox{
\includegraphics[width=9cm]{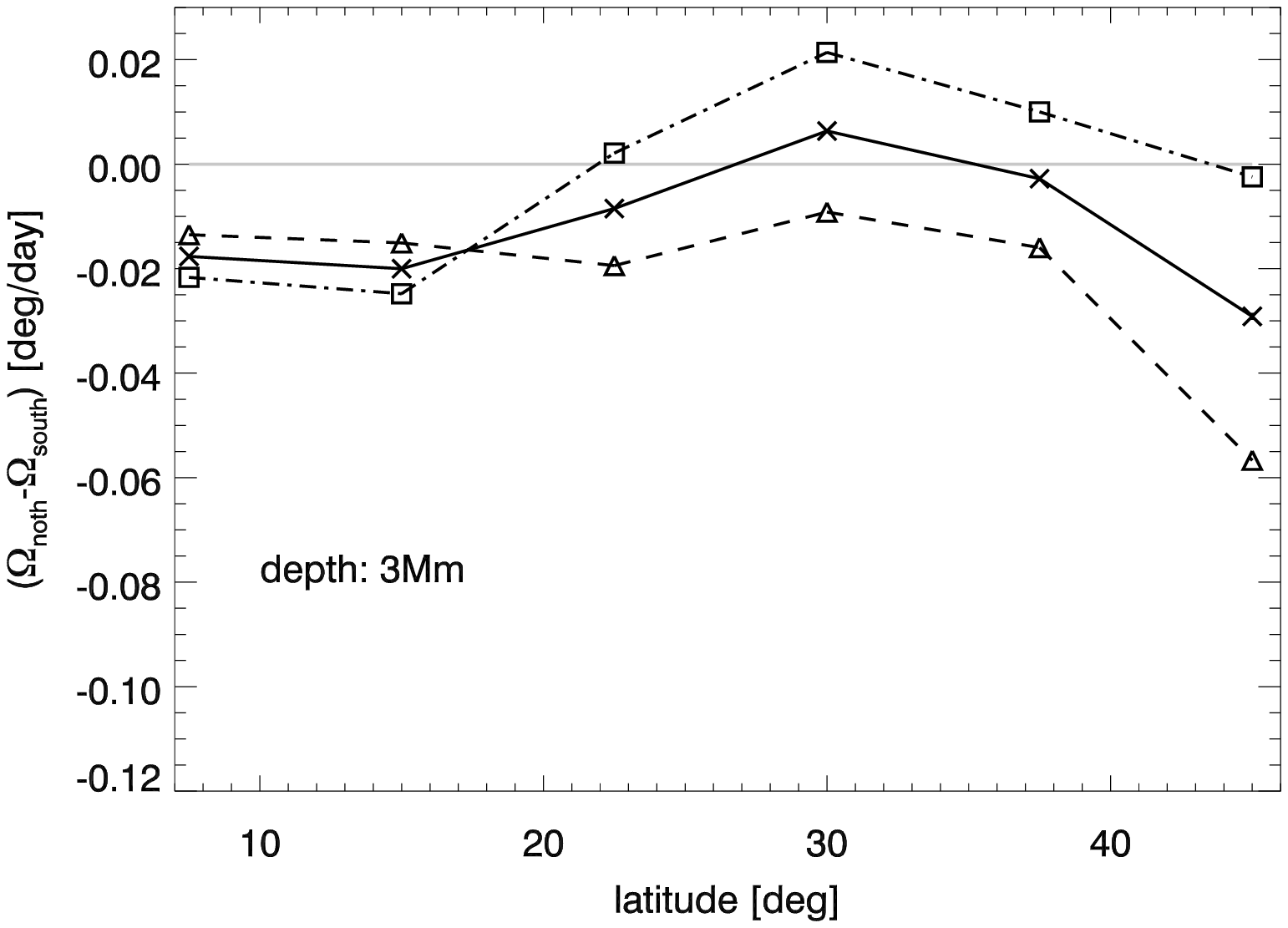}
\includegraphics[width=9cm]{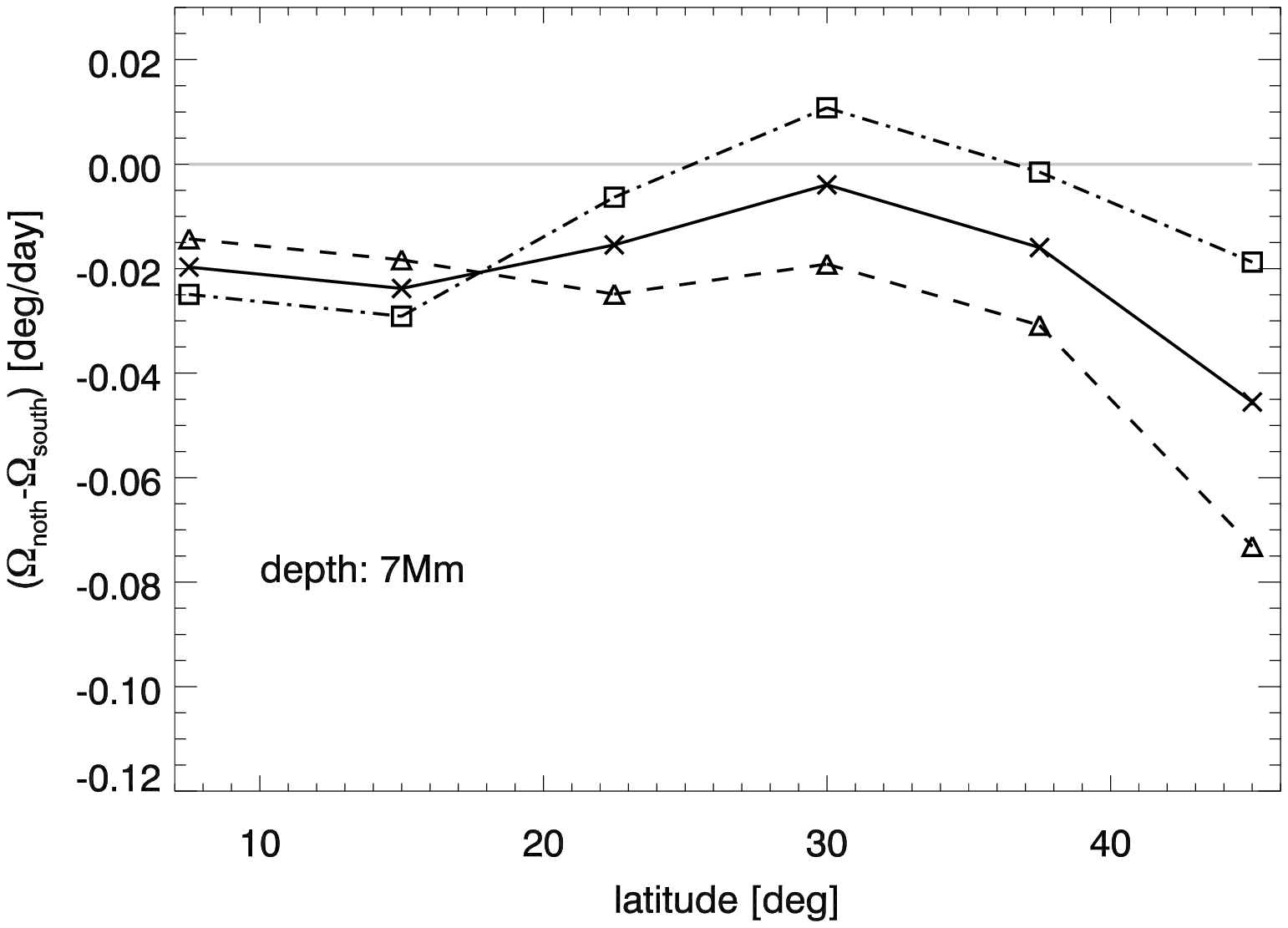}
\includegraphics[width=9cm]{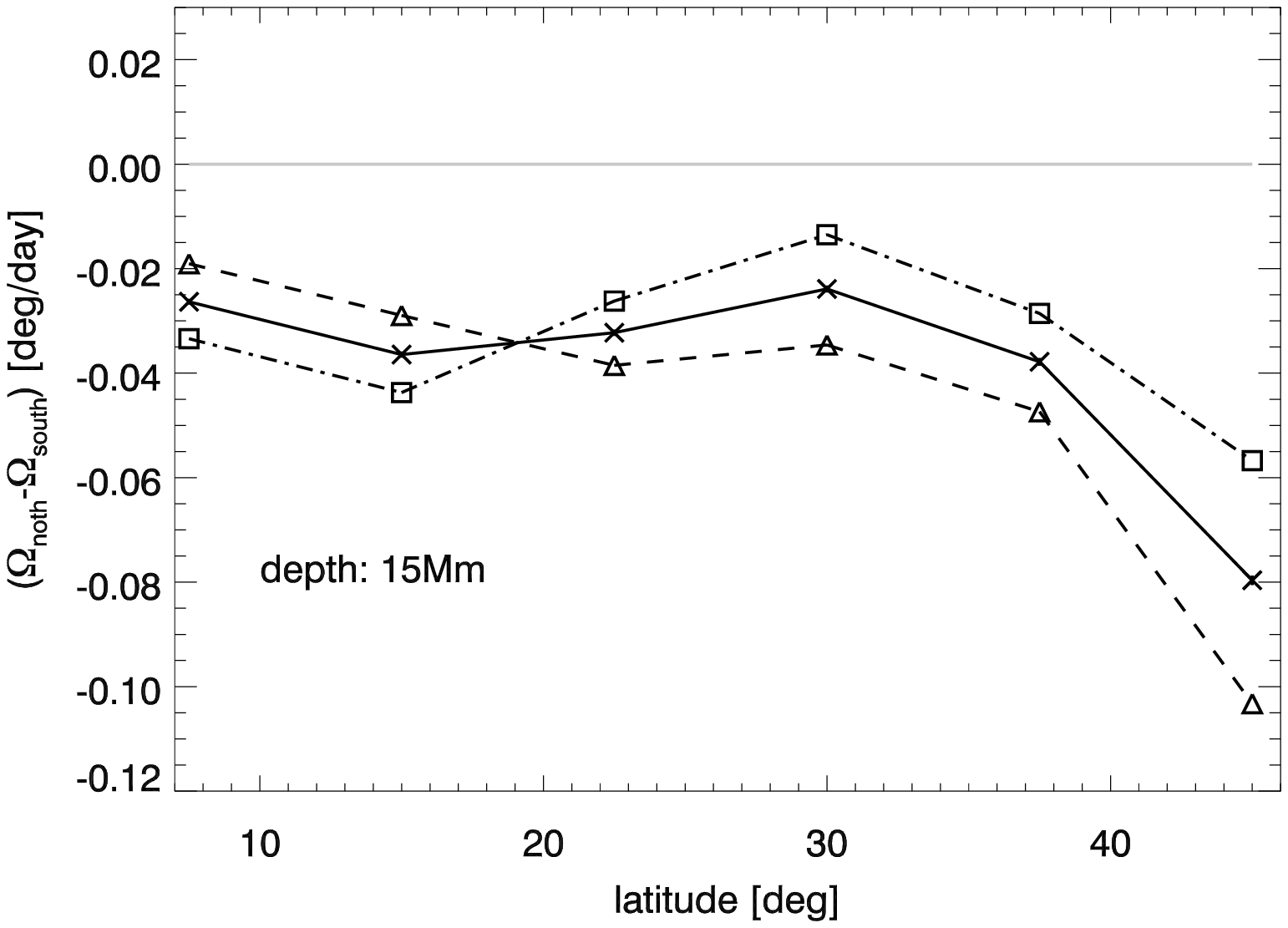}
}
}
\caption{Difference between the northern and southern averaged angular velocity of subsurface layers as a function of unsigned latitude and for depths 3Mm, 7Mm and 15Mm. Shown are the periods : August 2001 -- December 2006 (full lines), August 2001 -- March 2004 
(dashed), April 2004 -- December 2006 (dashed-dotted). The thin grey line shows the zero level. }
\label{n-s_rg}
\end{figure}

Figure \ref{n-s_rg} shows the difference between the northern and southern rotation velocities for three subsurface layers at depths of 3Mm, 7Mm, and 15Mm. As for the SBCS, subsurface rotation velocity is predominantly southern except at the intermediate latitudes of the upper subsurface layers during the low activity epoch. This southern dominance also characterised the photospheric magnetic activity during solar cycle 23 as shown from sunspot indicators (See Li et al. 2008 and references therein). Besides, Zaatri et al. (2006) have analysed the first three years of our GONG++ dataset and found that the zonal flow pattern and its north-south asymmetry are strongly related to the distribution of magnetic activity. Moreover, a clear increase of this asymmetry with depth is seen for the whole period. This confirms, for a longer period, the results by Zaatri et al. (2006) who found the same north-south asymmetry depth dependence for the zonal flow. Unlike the SBCS, the north-south asymmetry in the subsurface rotation velocity shows a slight increase with decreasing activity, at low latitudes. At latitudes higher than $20^{\circ}$ and close to solar cycle maximum, the subsurface asymmetry is slightly changing until around $40^{\circ}$. During the low activity epoch, the rotation velocity of the southern hemisphere gets very close to the northern velocity and even smaller around latitude $30^{\circ}$ for the upper subsurface layers.                          

Some hints of the north-south asymmetry are also given from the latitudinal gradient in Table 1 where the subsurface layers show more varying $B$ values with depth in the northern hemisphere than in the 
southern hemisphere for both activity levels. The SBCS have a more pronounced north-south 
variation of $B$ at low activity than at high activity. 

\section{Discussion and conclusion}
In this work, we compared the rotation of the corona as revealed by tracing its most profuse features, the SBCS,  and that of the uppermost layers of the convection zone. We report the following results  
 
(i) There is a difference of about $0.5^{\circ}$day$~^{-1}$ between the angular velocity of the SBCS and the upper considered subphotospheric layer at 3Mm below the surface. This difference decreases while considering deeper layers due to the increase of the angular velocity with depth at intermediate latitudes. At the highest latitudes of our considered range [$-52.5^{\circ},52.5^{\circ}$], the radial gradient of the rotation velocity decreases considerably and the difference between the angular velocities of the SBCS and the various subsurface layers becomes very similar. 

(ii) The temporal variation of the SBCS angular velocity during the declining phase of the solar cycle 23 shows an acceleration of about $0.005^{\circ}$day$~^{-1}$/month on average and reaches its maximum during the low activity period (end of cycle 23) in the southern hemisphere. It is not clear whether this acceleration is related to the chosen tracers or if it is the property of the corona and whether it is specific to the studied period. So far, there have been no studies of the coronal rotation using other activity features for the last years of solar cycle 23. However, results from other periods have reported a deceleration that makes the coronal rotation tend to that of the photosphere at solar minimum (e.g Giordano et al. 2008 for the period 1996-1997), whereas Mehta (2005) did not find any evidence of the coronal rotation variation during four consecutive solar cycles (19 to 22). The angular velocity does not show a significant temporal variation except a slight acceleration at high latitudes which corresponds to the torsional oscillation signal during the period 2001-2006. 

(iii) The latitudinal dependence of the angular velocity from the two measurements was studied through the differential rotation parameter $B$ as given by the most commonly used rotation law at intermediate latitudes $\theta$ ($\Omega(\theta)=A+B$sin$^2\theta$). We found a very similar latitudinal gradient $B$ of coronal and subphotospheric rotations, except that the unsigned value of $B$ slightly increases with depth. From this point of view, we do not see any particularity in the latitudinal behaviour of the coronal rotation as mentioned by several authors from long-term coronal features observations, notably, the widely mentioned existence of two different rotational modes with a more rigid rotation of the corona than for the photosphere at low latitudes (Altrock 2003). Clearly, the SBCS behave more like magnetic photospheric features which endorses the hypothesis of the formation of EUV bright points as being caused by the heating derived from the dissipation of electric currents that are formed in the solar atmosphere from the displacement of footpoints of magnetic flux tubes by the photospheric plasma motion (see Santos et al. 2008 and references therein).   
 
(iv) For both corona and subphotospheric layers, the southern hemisphere rotates predominantly faster than the northern hemisphere. The north-south velocity difference increases with depth. This result was already shown in Zaatri et al. (2006) and is now confirmed for a longer period including the very low activity period of cycle 23. 
Also, when the north-south asymmetry of the rotation velocity is lower than its temporal average for the  subsurface layers, it is higher for the SBCS and vice-versa.               
Clearly, the southern coronal hemisphere rotates faster than the northern hemisphere regardless of which hemisphere is more active. It would be interesting to see whether this persists on a time scale of multiple cycles as opposed to the relatively short time scale of our study.

\begin{acknowledgements}
A. Zaatri and M. Roth acknowledge support from the European Helio- and Asteroseismology
Network (HELAS) which is funded by the European Commission's Sixth Framework Programme.
The research of R. Braj\v sa leading to the results presented in this paper has received funding from the European
Community's Seventh Framework Programme (FP7/2007-2013) under grant agreement no. 218816.
The work on SBCS analysis was performed with the support of the Alexander von Humboldt Foundation and is
related to the SOHO-EIT Proposal Braj\v sa 206:"An analysis of the solar rotation velocity by tracing coronal
features" (http://umbra.nascom.nasa.gov/eit/proposals/) submitted in March 1999 by R. Braj\v sa, B. Vr\v snak, V. Ru\v zdjak, D. Ro\v sa, H. W\"ohl, and F. Clette. SOHO is a project of international
cooperation between ESA and NASA. We would like to thank the EIT team for developing and operating the
instrument. We thank F. Clette, J.-F. Hochedez, S.F. Gissot and J. de Patoul for
providing the EIT images used and helpful discussions. This work utilises data
obtained by the Global Oscillation Network Group (GONG) programme, managed by
the National Solar Observatory, which is operated by the Association of Universities
for Research in Astronomy (AURA), Inc. under a cooperative agreement with
the National Science Foundation. The data were acquired by instruments operated
by the Big Bear Solar Observatory, High Altitude Observatory, Learmonth Solar
Observatory, Udaipur Solar Observatory, Instituto de Astrofisica de Canarias, and
Cerro Tololo Interamerican Observatory.
\end{acknowledgements}

\begin{small}

 \end{small}

\begin{thebibliography}{}

\bibitem{altrock2003}
 Altrock, R. C. 2003, Solar Phys., 123, 23

 \bibitem{badalyan2006}
  Badalyan, O. G., S\'ykora, J. 2006, Advances in Space Research, 38, 906

 \bibitem{beck2000}
  Beck, J. G. 2000, Solar Phys., 191, 47

 \bibitem{Brajsa2001}	
  Braj\v sa, R., W\"ohl, H., Vr\v snak, B., Ru\v zdjak, V., Clette, F., Hochedez, J.-F. 2001, A\&A, 374, 309

 \bibitem{Brajsa2002}	
  Braj\v sa, R., W\"ohl, H., Vr\v snak, B., Ru\v zdjak, V., Clette, F., Hochedez, J.-F. 2002, A\&A, 392, 329

 \bibitem{Brajsa2004}
  Braj\v sa, R., W\"ohl, H., Vr\v snak, B., Ru\v zdjak, V., Clette, F., Hochedez, J.-F., Ro\v sa, D. 2004, AA, 414, 707

 \bibitem{Brajsa2006}
  Braj\v sa, R., Ru\v zdjak, D., W\"ohl, H. 2006, Solar Phys., 237, 365

 \bibitem{Brajsa2008}
  Braj\v sa, R., W\"ohl, H., Vr\v snak, B., Ru\v zdjak, V., Clette, F., Hochedez, J.-F., Verbanac, G., Skoki\' c, I., 
  Hanslmeier, A.  2008, Cent. Eur. Astrophys. Bull. 32, 165

 \bibitem{corbard2002}
  Corbard, T., Thompson, M. J. 2002, Solar Phys., 205, 211

 \bibitem{corbard2004}	
  Corbard, T., Toner, C., Hill, F., Hanna, K. D., Haber, D. A., Hindman, B. W., Bogart, R. S. 2003,
  in Proceedings of SOHO 12 / GONG+ 2002. Local and global helioseismology: the present and future, ESA SP-517, 255

 \bibitem{dsilva1996}
  D'Silva, S. 1996, ApJ, 462, 519

 \bibitem{giordano}
  Giordano, S., Mancuso, S. 2008, ApJ, 688, 656  

 \bibitem{hill1988}
  Hill, F. 1988, ApJ, 333, 996

 \bibitem{howard1980}
  Howard, R., LaBonte, B. J. 1980, ApJ, 239, 33

 \bibitem{howe2009}
  Howe, R. 2009, Living Reviews in Solar Physics, 6, lrsp-2009-1
  http://www.livingreviews.org/lrsp-2009-1

 \bibitem{Insley1995}
  Insley, J. E., Moore V., Harrison R. A. 1995, Solar Phys., 160, 1

 \bibitem{joshi07}
  Joshi, B., Pant, P., Manoharan, P.K., Pandey, K. 2007, ASP Conf. Ser., 368, 539

 \bibitem{karachik2006}
  Karachik, N., Pevtsov, A. A., Sattarov, I. 2006, ApJ, 642, 562

 \bibitem{koch1984}
  Koch, A. 1984, Solar Physics, 93, 53

 \bibitem{Komm2009}
  Komm, R., Howe, R., Hill, F., Gonz\'alez Hern\'andez, I. 2009, Solar Phys., 254, 1

 \bibitem{Li2008}
  Li, K.J., Gao, P.X., Zhan, L.S. 2009, Solar Phys., 254, 145 

 \bibitem{Mouradian}
  Mouradian, Z., Bocchia, R., Botton, C. 2002, A\&A, 394, 1103
 
 \bibitem{Mehta}
  Mehta, M. 2005, Bull. Astr. Soc. India, 33, 323

 \bibitem{obridko}
  Obridko, V. N.,Shelting, B. D. 2001, Sol. Phys., 201, 1

 \bibitem{Pres1999}
  Pres, P., Phillips, K. J. H. 1999, ApJ, 510, L73

 \bibitem{Ruz2004}
  Ru\v zdjak, D., Ru\v zdjak, V., Braj\v sa, R., W\"ohl, H. 2004, Solar Phys., 221, 225

 \bibitem{santos2008}
  Santos, J. C., B\"uchner, J., Madjarska, M. S., Alves, M. V. 2008, A\&A, 490, 345 

 \bibitem{Schroeter1985}
  Schr\"oter, E. H. 1985, Solar Phys., 100, 141

 \bibitem{zaatri2006}
  Zaatri, A., Komm, R., Gonz\'alez Hern\'andez, I., Howe, R., Corbard, T. 2006, Solar Phys., 236, 227

 \bibitem{Zaatri2009}
  Zaatri, A., Corbard, T. 2009, in proceedings of GONG 2008/SOHO XXI meeting ASP Conf. Ser., in press

 \end{thebibliography}
 \end{document}